\documentclass[12pt,fleqn]{article}
\usepackage{graphicx}
\begin{document}

\title{\bf {Modified Schr\"odinger equation, its analysis and experimental verification}}
\author{Isaac Shnaid}
\maketitle

\begin{abstract}
According to classical non-relativistic Schr\"odinger equation, any local perturbation of wave function instantaneously affects all infinite region, because this equation is of parabolic type, and its solutions demonstrate infinite speed of perturbations propagation. From physical point of view, this feature of Schr\"odinger equation  solutions is questionable. According to relativistic quantum mechanics, the perturbations  propagate with speed of light. However when appropriate mathematical procedures are applied to Dirac relativistic quantum equation with finite speed of the wave function perturbations propagation, only classical Schroedinger equation predicting infinite speed of the wave function perturbations propagation is obtained. Thus, in non-relativistic quantum mechanics the problem persists. 
In my work modified non-relativistic Schr\"odinger equation is formulated. It is also of parabolic type, but its solutions predict finite speed of the wave function perturbations propagation.  Properties of modified  Schr\"odinger equation solutions are studied.
I show that results of classical Davisson-Germer experiments with electron waves diffraction support developed theoretical concept of modified Schr\"odinger equation, and predict that speed of the wave function perturbations propagation has order of magnitude of speed of light. 
\end{abstract}



\section{Introduction}
We will start from classical time dependent Schr\"odinger equation 
\begin{equation}
i \hbar \frac {\partial \Psi}{\partial t}+ \frac{\hbar^2}{2 m} \nabla^2 \Psi - U \Psi=0
\label{eq:B1}
\end{equation}
and its corollary classical DeBroglie's  wave of a free material particle ($U=0,~E=const$) moving parallel axis $x_m$
\begin{equation}
\Psi=\exp[2 \pi i (-\nu t + k x_m)]
\label{eq:B2}
\end{equation} 
where $\hbar=h/2\pi,~i=\sqrt{-1}$, and $h,~\Psi,~t,~m,~U,~\nu=E/h,~k=p/h,~E,~p=m v,~v$ denote Planck's constant, wave function, time, mass, potential energy, frequency,  wave number, the particle energy, momentum and speed, respectively.

We suggest $U=0$ and analyse a system consisting of a material particle which initially does not move $E=0,~p=0$ with respect to a coordinate system $x,y,z$. It means that initially $\nu=0,~k=0$, DeBroglie's wave does not exist, and $\Psi=0$. Then at some moment of time $t_0=0$ the particle starts moving, and becomes $E \neq 0,~p \neq 0$, thus, creating a non-zero DeBroglie's wave $\nu \neq 0,~k \neq 0$ and a non-zero $\Psi \neq 0$ local perturbation of the wave function.  In this case, according to equation (\ref{eq:B1}),  all  infinite region instantaneously becomes perturbed: everywhere instantaneously becomes $\Psi \neq 0$,  because this equation is of parabolic type and therefore predicts infinite speed of the perturbations propagation. Also instantaneously infinite DeBroglie's wave (\ref{eq:B2}) appears in the region. From physical point of view, such mathematical behaviour of classical time dependent Schr\"odinger equation solutions is questionable. 

According to relativistic quantum mechanics, the perturbations propagate with speed of light. However when appropriate mathematical procedures are applied to Dirac relativistic quantum equation with finite speed of the wave function perturbations propagation, only classical Schroedinger equation predicting infinite speed of the wave function perturbations propagation is obtained. Thus, in non-relativistic quantum mechanics the problem persists. 

I suggest that in non-relativistic case speed of the perturbations propagation is also finite and is equal to speed of light. In my works \cite{sh1}-\cite{sh3} were derived and studied parabolic partial differential equations with solutions predicting finite speed of the perturbations propagation. These findings I apply now in order to formulate and analyse modified time dependent Schr\"odinger equation whose solutions display finite speed of the wave function perturbations propagation. As one of results I prove that electron wavelength is smaller than determined from classical equations, and find that Davisson and Germer experiment confirms this prediction.

\section {Theory of the modified time dependent Schr\"odinger equation}

\subsection{The wave function perturbations propagation}

We assume that initially zero wave function $\Psi=0$ exists in infinite region. At  moment of time $t_0=0$ a non-zero perturbation of the wave function $\Psi \neq 0$ is locally introduced, and finite speed of the perturbations propagation is assumed. Therefore a non-perturbed $\Psi=0$ and a perturbed $\Psi \neq 0$ subregions appear in the region. Let $S_P$ be a surface separating the perturbed subregion from the non-perturbed one. According to our basic assumption, the perturbed subregion gradually propagates into the non-perturbed subregion, and any point $E$ of the surface $S_P$ moves  with finite normal speed $v_P$ which is speed of the perturbations propagation. As speed of the perturbations propagation $v_P$ cannot be lower than  the particle speed  $v$, therefore $v_P$ is equal to speed of light. 

Let perturbation traveltime $t_P(x,y,z)$ \cite{sh1}-\cite{sh3} be a time moment when the perturbation reached a given point $M(x,y,z)$. It means that the boundary surface $S_P$, separating the perturbed from the non-perturbed subregion, is a surface of constant perturbation traveltime  $t_P(x,y,z)=const$. From definition of perturbation traveltime follows

\begin{equation}
|\nabla t_P|=\frac {1}{v_P}
\label{eq:A1}
\end{equation}
\begin{equation} 
\bigg ( \frac {\partial t_P}{\partial x}\bigg )^2+\bigg ( \frac {\partial t_P}{\partial y}\bigg )^2+\bigg ( \frac {\partial t_P}{\partial z}\bigg )^2=\frac{1}{v_P^2}
\label{eq:A2}
\end{equation}

The non-linear governing equation (\ref{eq:A2}) is similar to the classical eikonal equation \cite{sh4}, \cite{sh5}. Its primary wave solution $t_P(x,y,z)$ defines perturbation traveltime  and satisfies initial condition 
\begin{equation}
t_{0P}(x,y,z)=t_0=0
\label{eq:A3}
\end{equation}
where $t_{0P}(x,y,z)$ and $t_0$ denote initial value of traveltime and time moment when the perturbation was introduced.

\subsection{The local time concept and the modified Schr\"odinger equation}

For any point $M(x,y,z)$ of the region, can be introduced local time $\vartheta(x,y,z,t)$ \cite {sh1}-\cite{sh3} counted from a moment when the perturbation reached this point   
\begin{equation}
\vartheta(x,y,z,t)=t-t_P(x,y,z)
\label{eq:A4}
\end{equation}

Naturally, there can be three characteristic cases:
\begin{enumerate}
\item $\vartheta=t-t_P<0$, the perturbation has not reached the point $M$. Thus, the point belongs to the non-perturbed subregion.
\item $\vartheta=t-t_P=0$, the perturbation has reached the point $M$. Now the point $M$ is located on the border surface $S_P$, separating the perturbed and the non-perturbed subregions.
\item $\vartheta=t-t_P>0$, the point $M$ is located inside the finite perturbed subregion with moving border $S_P$. All further considerations are related to this subregion where $\vartheta > 0,~\Psi \neq 0,~\nu \neq 0,~k \neq 0$. Obviously, classical time dependent Schr\"odinger equation does not describe such situation, and a new quantum equation must be derived.
\end{enumerate}

According to the local time concept formulated earlier \cite{sh1}-\cite{sh3}, parabolic type partial differential equations predicting finite speed  of the perturbations propagation and the respective equations with infinite speed  of the perturbations propagation are identical, if independent variables  $\vartheta=t-t_P,~x,~y,~z$ are used. When speed of the perturbations propagation is infinite $v_P \rightarrow \infty$, in these equations local time is identical to global time $\vartheta=t$ because $ t_P=0 $. The same equations describe the finite $v_P$ case  when $t_P > 0,~ 0 < \vartheta < t$. All equations written in such universal form we call \emph {modified  equations}.

Application of local time concept leads to modified time dependent Schr\"odinger equation
\begin{equation}
i \hbar \frac {\partial \Psi}{\partial \vartheta}+ \frac{\hbar^2}{2 m} \nabla^2 \Psi - U \Psi=0
\label{eq:A5}
\end{equation}
and its corollary modified DeBroglie's wave equation
\begin{equation}
\Psi=\exp[2 \pi i (-\nu \vartheta + k x_m)]
\label{eq:A6}
\end{equation}

We will prove that solutions of modified time dependent Schr\"odinger equation predict finite speed of the perturbations propagation. 
 
\section{Analysis of the modified Schr\"odinger equation}

\subsection{Comparison of modified equation with classical Schr\"odinger equation}

Modified time dependent Schr\"odinger equation (\ref{eq:A5}) is of the parabolic type as classical Schr\"odinger equation. In a classical case of infinite speed of the perturbations propagation $v_P \rightarrow \infty$ there is $t_P=0,~\vartheta=t$, and modified equations (\ref{eq:A5}) and (\ref{eq:A6}) become identical to classical equations (\ref{eq:B1}) and (\ref{eq:B2}), respectively.

Solutions of classical time dependent Schr\"odinger parabolic type equation (\ref{eq:B1}) predict that any local perturbation introduced at initial  moment of time $t_0=0$, instantaneously affects all infinite space domain. The modified time dependent Schr\"odinger parabolic type equation (\ref{eq:A5}) uses local time $\vartheta$ instead of global time $t$ as an independent variable. In this case, initial perturbation introduced at initial local time moment $\vartheta_0=t_0-t_{0P}=0$,  affects an arbitrary point $A(x_A,y_A,z_A)$ of the space domain at the same local time value $\vartheta_A=t_A-t_P(x_A,y_A,z_A)=0$.  Therefore perturbation arrives to an arbitrary point $A(x_A,y_A,z_A)$ at global time moment $t_A=t_P(x_A,y_A,z_A)$, i.e. with global time delay. So the modified Schr\"odinger equation predicts finite speed of the perturbations propagation, and the perturbations have a wave behaviour. These waves do not reflect and interfere, because only solutions of eikonal  equation (\ref{eq:A2}) corresponding to primary perturbation waves propagating in the non-perturbed subregion, have physical meaning. 

Let  $\Psi(x,y,z,t)$ be a solution of the classical  Schr\"odinger equation (\ref{eq:B1}), satisfying certain boundary and initial conditions. For the same boundary and initial conditions, a solution of the modified Schr\"odinger equation (\ref{eq:A5}) is the same function but with local time as an argument $\Psi(x,y,z,\vartheta),~\vartheta>0$. For an arbitrary point $A$  with coordinates $x_A,~y_A,~z_A$ and perturbation traveltime value $t_P(x_A,y_A,z_A)$, located inside the perturbed subregion, the difference of solutions of classical and modified  Schr\"odinger equations is 
\begin{equation}
\Delta \Psi=\Psi(x_A,y_A,z_A,t)-\Psi(x_A,y_A,z_A,\vartheta=t-t_P) \approx \frac {\partial \Psi}{\partial t} ~ t_P
\label{eq:A8}
\end{equation}
Absolute value of this difference is small for small values of perturbation traveltime $t_P$ and/or slow processes, when absolute value of $\frac {\partial \Psi}{\partial t}$ is small. In this case, solutions of classical time dependent Schr\"odinger equation are accurate enough.

\subsection{Time independent modified equation}

By using a standard presentation of $\Psi=\psi \exp (-2 \pi i \nu t)$ for $E=const$ \cite{sh6}-\cite{sh8} but with local time $\vartheta$ as an argument  $\Psi=\psi \exp (-2 \pi i \nu \vartheta)$ we obtain time independent version of modified Schr\"odinger equation (\ref{eq:A5})

\begin{equation}
\frac{\hbar^2}{2 m} \nabla^2 \psi +(E - U) \psi=0
\label{eq:A7}
\end{equation}

We see that time independent version of modified Schr\"odinger equation (\ref{eq:A7}) is identical to classical time independent Schr\"odinger equation. Therefore all solutions of modified time independent Schr\"odinger equation formally are identical to solutions of classical equation. However there is also an essential difference, because in the classical case multiplier $\exp (-2 \pi i \nu t)$ is only function of time $t$, while in the case of modified Schr\"odinger equation multiplier $\exp (-2 \pi i \nu \vartheta)=\exp [-2 \pi i \nu (t-t_P(x,y,z))]$ is a function of time and space coordinates which exists only in perturbed subregion where $t>t_P(x,y,z)$. Therefore any solution of modified time independent equation has physical meaning only in finite perturbed subregion with moving border and inside this subregion wave function can be normalized. 

\section{Waves of matter according to modified Schr\"odinger equation}

In a general case function $\psi$ is complex 

\begin{equation}
\psi(x,y,z)=|\psi(x,y,z)|~ exp [2\pi i \phi(x,y,z)] 
\label{eq:A7a}
\end{equation}
where $2\pi \phi(x,y,z)$ is argument of $\psi$. So wave function $\Psi$ for modified Schr\"odinger equation (\ref{eq:A5}) can be written as

\begin{equation}
\Psi=\psi \exp (-2 \pi i \nu \vartheta)=|\psi| \exp [2 \pi i( -\nu t+\Phi(x,y,z))] 
\label{eq:A7b}
\end{equation}

\begin{equation}
\Phi(x,y,z)=\nu t_P(x,y,z)+\phi(x,y,z)
\label{eq:A7c}
\end{equation}

As $k_l=|\nabla \Phi(x,y,z)|$ is wave number of matter wave according to modified Schr\"odinger equation, $k=|\nabla \phi(x,y,z)|$ is wave number of matter wave according to classical Schr\"odinger equation, and $|\nabla t_P|=1/v_P$, the following expression can be obtained from formula (\ref{eq:A7c})

\begin{equation}
k_l^2=k^2+ \frac{\nu^2}{v_P^2}+ \frac{2 \nu k}{v_P} \cos \alpha
\label{eq:A7d}
\end{equation}
where $\alpha$ denotes an angle between vectors $\nabla \phi$ and $\nabla t_P$. 

For $\cos\alpha \geq -\nu/2 k v_P$ there is $k_l \geq k,~\lambda_l \leq \lambda$; for $\cos\alpha < -\nu/2 k v_P$ there is $k_l<k,~\lambda_l>\lambda$; and for $\cos\alpha = -\nu/2 k v_P$ will be $k_l=k,~\lambda_l=\lambda$. In these formulae  $\lambda_l=1/k_l,~\lambda=1/k $ denote wavelengths of matter wave defined by modified and classical Schr\"odinger equations, respectively. 

For a characteristic case of $\alpha=0$ we obtain from expression (\ref{eq:A7d})

\begin{equation}
k_l=k+ \frac{\nu}{v_P}
\label{eq:A7e}
\end{equation}

Formula (\ref{eq:A7e}) proves that if $\alpha=0$, then modified  wave number $k_l$ is higher than classical wave number $k$. Therefore modified wavelength $\lambda_l=1/k_l$ is lower than classical wavelength $\lambda=1/k$.

Let $v_{ph.l}=\nu/k_l,~v_{gr.l}=d\nu/dk_l,~v_{ph}=\nu/k,~v_{gr}=d\nu/dk$ be modified wave phase velocity, modified wave group velocity, classical wave phase velocity and classical wave group velocity, respectively. In the case of $\alpha=0$ phase and group velocities of matter waves determined from modified formula (\ref{eq:A7e}) are smaller than defined by classical Schr\"odinger equation

\begin{equation}
\frac{1}{v_{ph.l}}=\frac{1}{v_{ph}}+\frac{1}{v_P}
\label{eq:A7g}
\end{equation}
\begin{equation}
\frac{1}{v_{gr.l}}=\frac{1}{v_{gr}}+\frac{1}{v_P}
\label{eq:A7h}
\end{equation}

If  a free material particle ($U=0,~\nu=m v^2/2h,~k=m v/h$) initially was at $x_m=0$ and moves parallel axis $x_m>0$, then perturbation traveltime and local time are defined as $t_P=x_m/v_P,~\vartheta=t-x_m/v_P$. Thus, modified DeBroglie's wave equation (\ref{eq:A6}) becomes

\begin{equation}
\Psi= \exp[2 \pi i (-\nu t + k_l x_m)]
\label{eq:A9}
\end{equation}
where
\begin{equation}
k_l=k+\frac{\nu}{v_P}=\frac {m v}{h} \bigg (1+\frac{v}{2 v_P} \bigg)
\label{eq:A9a}
\end{equation} 
is modified DeBroglie's wave number which is higher than classical wave number $k=mv/h$. It means that modified DeBroglie's wavelength 
$\lambda_l=1/k_l$ is lower than classical DeBroglie's wavelength $\lambda=1/k$. Expression (\ref{eq:A9a}) is a particular case of (\ref{eq:A7e}).

The modified DeBroglie's wave phase $v_{ph.l}$ and group $v_{gr.l}$ velocities are smaller than respective classical values \cite{sh6}-\cite{sh8} $v_{ph}=v/2$ and $v_{gr}=v$ 

\begin{equation}
v_{ph.l}=\frac{\nu}{k_l}=\frac{v_{ph}}{1+v_{ph}/v_P}=\frac{v}{2~(1+v/2v_P)}
\label{eq:A10}
\end{equation}
\begin{equation}
v_{gr.l}=\frac {d\nu}{dv} \frac{dv}{d k_l}=\frac {v_{gr}} {1+v_{gr}/v_P}= \frac{v}{1+v/v_P}
\label{eq:A11}
\end{equation}
Formulae (\ref{eq:A10}) and (\ref{eq:A11}) are similar to (\ref{eq:A7g}) and (\ref{eq:A7h}), respectively.

\section{Modified Schr\"odinger equation and experiments with electron waves diffraction}

Nobel lecture of C.J. Davisson \cite{sh10} includes results of classical Davisson and Germer experiment with electron waves diffraction on a crystal of nickel.  The lecture presents experimental values of electron wave length $\lambda_{exp}$ vs. $V^{-0.5}$, where $V$ is electron accelerating voltage. These data can be used for experimental check of formula (\ref{eq:A9a}) because $k_{exp}=1/\lambda_{exp}$ and $v=(2 e V/m)^{0.5}$, where $e,~k_{exp}$ denote electric charge of electron and experimental value of electron wave number, respectively.

\begin{figure}
\center
\includegraphics[width=12cm, height=10cm]{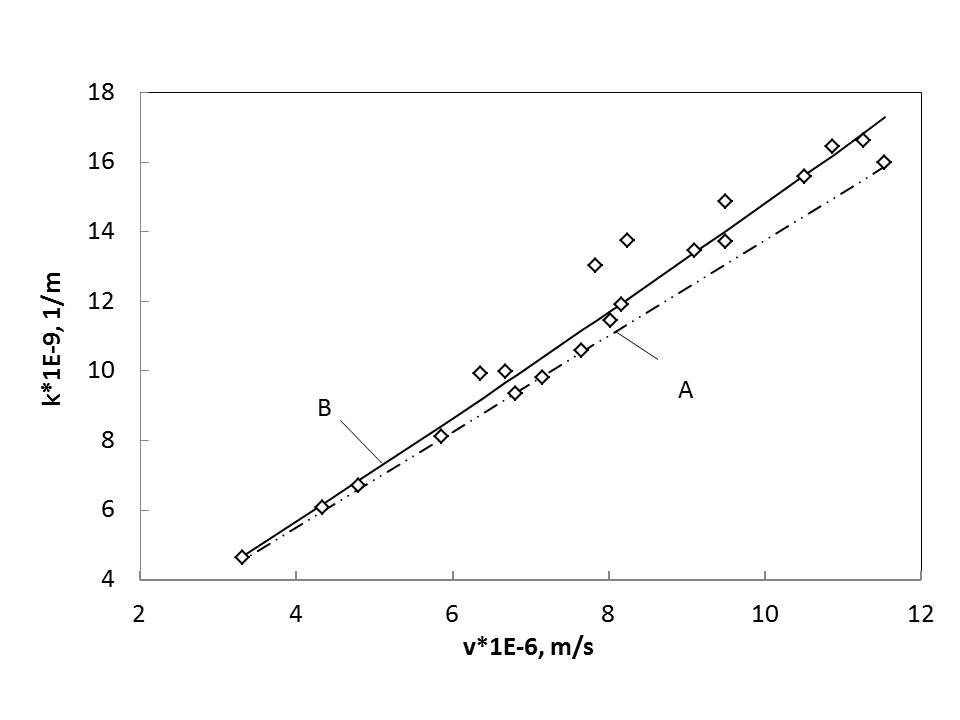}
\caption{Electron wave number vs. electron speed in Davisson and Jermer experiment (diamonds - experimental data from \cite{sh10}; A - theoretical curve for classical DeBroglie's wave with infinite speed of the perturbations propagation; B - best fit curve for modified DeBroglie's wave with finite speed of the perturbations propagation).}
\end{figure} 

Fig. 1 shows experimental points (diamonds) with values of $k_{exp}$ and $v$ corresponding to respective values  of $\lambda_{exp}$ and $V$ in lecture \cite{sh10}. Line A is classical $v_P=\infty$ DeBroglie's electron wave number $k=m v/h$ vs. electron speed $v$. Though a part of experimental points are close enough to A $k_{exp} \approx k$, for most of experimental points there is significantly $k_{exp} > k$, and are missing points where substantially $k_{exp} < k$.

Line B in Fig. 1 is best fit (least  squares) approximation for all experimental points, obtained by variation of $v_P$ in formula (\ref{eq:A9a}) for modified wave number $k_l$. For line B variance is $4.9*10^{17}~ 1/m^2$, and for line A variance is significantly bigger $11.2*10^{17}~ 1/m^2$. Line B corresponds to speed of the perturbations propagation equal $v_P=1.3*10^8$ m/s, which is 44\% of speed of light. Author of paper \cite{sh10} does not present experimental errors. Because of this, we are not able to find why experimental value of speed of the perturbations propagation differs from our theoretical prediction. Most probably it is the result of experimental errors. Anyway obtained experimental value of speed of the wave function perturbations propagation has order of magnitude of speed of light.

So analysed experimental data 
\begin{enumerate}
\item confirms correctness of formula (\ref{eq:A9a}); 
\item supports hypothesis of finite speed of the wave function perturbations propagation, and supplies its experimental value of $v_P=1.3*10^8$ m/s, which is 44\% of speed of light;
\item proves that modified Schr\"odinger equation is physically sound.
\end{enumerate}

\section{Conclusions}

\hspace{5mm} Derived modified time dependent Schr\"odinger equation is of parabolic type, as the classical equation. Its solutions predict finite speed of the wave function perturbations propagation.

Properties of modified  Schr\"odinger equation solutions are studied. I prove that classical time dependent Schr\"odinger equation is a particular case of modified equation, and time independent versions of modified and classical Schr\"odinger equations are identical.

Results of classical Davisson-Germer experiments with electron waves diffraction support developed theoretical concept of modified Schr\"odinger equation, and predict that speed of the wave function perturbations propagation has order of magnitude of speed of light.

\end{document}